# Field induced large magnetocaloric effect and magnetoresistance in ErNiSi

Sachin Gupta,[1] R. Rawat[2] and K. G. Suresh[1,*]

[1]*Department of Physics, Indian Institute of Technology Bombay, Mumbai-400076, India*

[2]*UGC-DAE Consortium for Scientific Research, Indore-452017, India*

**Abstract**

Large magnetocaloric effect (MCE) and magnetoresistance (MR) together with negligible hysteresis loss has been observed in ErNiSi compound, which undergoes metamagnetic transition at low temperatures. Magnetization, heat capacity and resistivity measurements confirm the metamagnetic transition. The maximum value of isothermal entropy change ($\Delta S_M$) and MR for a field change of 50 kOe are found to be 19.1 J/kg K and -34 %. Large MCE with negligible magnetic hysteresis loss could make this material promising for low temperature magnetic refrigeration.

PACS: 75.30.Kz, 75.47.De, 75.30.Sg

Keywords: metamagnetic transition, magnetoresistance, magnetocaloric effect

*corresponding author (email: suresh@phy.iitb.ac.in)

Magnetocaloric effect (MCE) is the thermal response of a magnetic material on the application/ removal of the magnetic field. It is characterized in terms of isothermal magnetic entropy change ($\Delta S_M$) and/or adiabatic temperature change ($\Delta T_{ad}$). Magnetic refrigeration based on MCE is an alternative to the conventional gas cycle refrigeration technology and has many advantages.[1] Considerable research in the past has been devoted to room temperature magnetic refrigeration, but some work has also been carried out for sub room temperature regime, which can be utilized in applications such as liquefaction of $H_2$ (20 K) and He (4 K). Zhang et al.[2] reported that significant cost reduction can be achieved with the utilization of magnetic hydrogen liquefier compared to conventional liquefiers. This technology could make hydrogen to be competitive as an alternative fuel.[3] Magnetic refrigeration at liquid He temperature is of interest in space applications too.[1] Chen et al.[4] have recently reported the development of an Active Magnetic Regenerative Refrigeration (AMRR) for space application. The technology employs a reversible circulator to circulate He bi-directionally through magnetic regenerators and can continuously provide remote distribute /distributed cooling near 2 K and reject heat beyond 15 K.[4] Large MCE has been reported in several ferromagnetic/antiferromagnetic materials at low temperatures.[5-9] A large MCE in a material arises due to the change in the magnetic moment configuration on the application of an external field. This change may also affect the scattering of conduction electrons and may result in large magnetoresistance (MR).[10]

It has been observed that RTX (R=rare earth, T= transition metal, X= p-block element) compounds show many interesting physical properties. Some of these compounds show large MCE and MR at low temperatures and could be promising for low temperature magnetic refrigeration and magnetic sensors.[11-15] A report on ErNiSi was published by Szytuła et al.,[16] in which the authors carried out neutron diffraction and magnetometric measurements. The neutron report[16] shows a collinear antiferromagnetic structure with a propagation vector of (0,1/2,1/4) at 1.5 K, which transforms into a sine-modulated magnetic structure with a propagation vector of (0.197, 0.515, 0.154) at 2.2 K. It is worth mentioning that ErNiSi shows a commensurate magnetic structure at low temperatures, which changes to an incommensurate structure at higher temperature (i.e. near the Neel temperature, $T_N$).[16] In this paper, we study ErNiSi by means of magnetic, magnetocaloric and magneto-transport measurements and try to correlate magnetic, magnetocaloric and magnetotransport properties to get a complete picture of the underlying magnetism in this compound and also to know its application potential.

The polycrystalline ErNiSi sample was synthesized using arc melt technique. The starting elements (Er, Ni, Si) with a purity at least 99.9 % were melted in a water cooled copper hearth using titanium as oxygen getter under argon atmosphere. The formed ingot was flipped up and melted several times for better homogeneity. As-cast sample was sealed in evacuated quartz tube and annealed for 100 hrs. at 800 °C followed by furnace cooling. X-ray powder diffraction (XRPD) pattern at room temperature was obtained from X'Pert Pro diffractometer (CuKα radiation). Magnetization, M(T, H) and the heat capacity, C(T) measurements were carried out on Quantum Design, Physical Property Measurement System (PPMS-6500). The magneto-transport measurements were carried out in a home made system, applying a current of 100 mA.

The Rietveld analysis of room temperature XRD pattern shows TiNiSi type orthorhombic crystal structure with space group Pnma. The lattice parameters obtained from the refinement are a= 686 pm, b=414 pm, c=716 pm, which are close to the reported values.[16]

Fig. 1(a) shows the temperature dependence of dc magnetic susceptibility ($\chi$) in zero-field cooled (ZFC) and field-cooled (FC) modes in 500 Oe. The compound shows a single peak characterizing antiferromagnetic ordering ($T_N$=3.2 K). In the paramagnetic region, it follows the Curie-Weiss law $\chi^{-1}$=(T-$\theta_{cw}$)/$C_m$ (where $\theta_{cw}$ is Curie-Weiss temperature and $C_m$ is molecular Curie constant). The value of the effective magnetic moment ($\mu_{eff}$) is 9.6 $\mu_B$/Er$^{3+}$ and $\theta_{cw}$ is close to 0 K. The $\mu_{eff}$ is very close to the expected moment of Er$^{3+}$ ion ($\mu_{th}$=9.59 $\mu_B$/Er$^{3+}$). The zero value of $\theta_{cw}$ can be explained on the basis of the neutron diffraction data[16], which show ferromagnetically coupled in-plane moments having antiferromagnetic inter-planar interaction. Due to these competing interactions, the net exchange becomes negligible, resulting in zero $\theta_{cw}$ value. As can be seen from Fig. 1(a), there is no difference between ZFC and FC curves, which indicates that the compound shows reversible behaviour and no thermal hysteresis. Fig. 1(b) shows the $\chi$ vs. T plots at different fields on a semi-log scale. One can see that the peak in $\chi$ vs. T plot broadens with increase in field. For fields ≥20 kOe, the plot resembles that of a ferro-para magnetic transition. The change in the $\chi$ data around $T_N$ suggests field induced metamagnetic transition.

Fig. 2 shows the field dependence of magnetization isotherms in the temperature range of 2- 60 K with a temperature interval of 2 K. Initially, the magnetization increases linearly with fields upto 10 kOe and above this, the magnetization increases sharply up to 20 kOe followed by a curvature at higher fields. The compound shows a tendency of saturation at 50 kOe. Linear increase in magnetization for low fields reveals the antiferromagnetic state, which transforms into a ferromagnetic state at higher fields due to the field induced metamagnetic transition. At 2, 4 and 6 K, the magnetization was recorded with increasing and decreasing field, which shows negligible hysteresis (see Fig. 2). This illustrates the reversible behavior due to the soft magnetic nature of the compound. Another point to be noted from Fig. 2 is that there is large magnetization change at different temperatures near the ordering temperature, which can result in large isothermal magnetic entropy change. Soft magnetic nature and the large magnetization difference are positive aspects for a potential magnetic refrigerant.

The temperature dependence of heat capacity at 0 and 20 kOe is shown in Fig. 3. The zero field heat capacity shows a λ-type peak at the onset of antiferromagnetic ordering, which suggests the second order nature of the magnetic phase transition in this compound. Upon application of a field, the peak in the heat capacity data broadens and shifts to higher temperatures, which is a feature characteristic of a field-induced metamagnetic transition. The moments are randomly oriented in the paramagnetic state and are aligned in ordered state. When the field is applied to the material, the ordering of random spins takes place along the field direction, which adds an additional ordering and results in the broadening of the heat capacity peak. Above $T_N$, the heat capacity shows a hump, which might be attributed to Schottky anomaly.

The MCE in ErNiSi has been estimated from the magnetization as well as the heat capacity data. The $\Delta S_M$ has been determined from the magnetization data employing the Maxwell's relation, $\Delta S_M = \int_0^H \left(\frac{\partial M}{\partial T}\right)_H dH$, where M is the magnetization and H is the applied field.[1] The $\Delta S_M$ and $\Delta T_{ad}$ have been estimated from heat capacity data using following equations[1]:

$$\Delta S_M(T,H) = \int_0^T \frac{C(T',H) - C(T',0)}{T'} dT' \qquad (1)$$

$$\Delta T_{ad}(T)_{\Delta H} \cong [T(S)_{H_f} - T(S)_{H_i}]_S \qquad (2)$$

The temperature dependence -$\Delta S_M$ estimated from the M-H-T and the C-H-T data is shown in Fig. 4. The inset in Fig 4(b) shows the temperature dependence of $\Delta T_{ad}$ at 20 kOe. The compound shows a peak around the ordering temperature in all the fields. The peak height and the width increase with increase in field. Aniferromagnetic materials generally show negative MCE while ferromagnetic materials show positive MCE. The positive MCE in this compound confirms that the field induced metamagnetic transition changes its state from antiferromagnetic to ferromagnetic. As can be seen from Fig. 4(a), MCE shows a sharp change at 20 kOe and is consistent with the sharp increase in the magnetization as the field is changed from 10 to 20 kOe. The small difference in the value of $\Delta S_M$ estimated from the two methods (for 20 kOe) can be seen, which may arise due to the different temperature steps in the two measurements. The compound shows large MCE associated with the metamagnetic transition as in a field of 50 kOe, the MCE reaches 19.1 J/kg K. This is larger than that of many potential magnetic refrigerants such as $ErRu_2Si_2$ (17.6 J.kg K),[9] HoAgGa (16 J/kg K),[11] $Dy_{0.9}Tm_{0.1}Ni_2B_2C$ (14.7 J/kg K),[17] $Er_4NiCd$ (18.3 J/kg K)[18], working in the same temperature range and subjected to same field change. The value of $\Delta T_{ad}$ is 2.5 K for the field of 20 kOe, which is also comparable to that of many intermetallic compounds.

In view of the large MCE seen in ErNiSi, we probed the MR behavior by measuring the electrical resistivity in different fields and is shown in Fig. 5(a). The main panel of Fig. 5(a) shows the zero field resistivity. The resistivity shows a peak at 3.2 K, which corresponds to the antiferromagnetic ordering temperature. At higher temperatures, the resistivity is almost linear and increases with temperature, revealing the metallic nature. The inset in Fig. 5(a) displays the resistivity in different fields. One can see that the peak gets suppressed and eventually disappears for the field of 50 kOe. The reduction in the resistivity near $T_N$ on the application of field occurs because of the loss of spin disorder in resistivity contribution. The application of the field reorients the moments from an antiparallel arrangement (high resistance state) to the parallel arrangement (low resistance state) and hence results in large magnetoresistance, which can be seen in Fig. 5(b). The main panel of Fig. 5(b) shows the temperature dependence of MR at 20 and 50 kOe. The magnitude of MR increases with decrease in temperature and becomes a maximum at 3.2 K ($T_N$). The maximum MR for the field change of 20 kOe becomes ~27 % which increases to ~34 % for 50 kOe. The maximum MR in ErNiSi is comparable to or larger than that of some intermetallic compounds such as $TbPd_3$ (~30 % at 70 kOe)[19] and NdCuSi (36%)[15], working in the same temperature range. It is worth mentioning here that the maximum MR obtained in ErNiSi is also larger than that of a few intermetallic compounds with metamagnetic transition and high ordering temperatures, such as $Gd_5(Si_{1.8}Ge_{2.2})$ (20%),[20] $SmMn_2Ge_2$ (8%),[21] $Ce(FeRu)_2$ (20%),[22] $Gd_5(Si_2Ge_2)$ (26%)[23] and $Gd_2In$ (29%)[24]. The field dependence of MR in increasing and decreasing fields at different temperatures is shown in the inset of Fig 5(b). It is clear from the inset that there is no hysteresis in the MR data. One can see that the magnitude of MR increases with decrease in temperature and increases with field. It attains the maximum at 3 K (i.e. near $T_N$). The MR shows a tendency of saturation at low temperatures and higher fields, as seen in the magnetization data.

ErNiSi crystallizes in the orthorhombic crystal structure and shows antiferromagnetic ordering. The magnetization, heat capacity, magnetocaloric and magnetoresistance measurements confirm field induced ferromagnetic state in this compound. The MCE has been estimated in terms of $\Delta S_M$ and $\Delta T_{ad}$ from the magnetization and heat capacity data. The compound shows large MCE and MR which is associated with the metamagnetic transition. The large MCE with negligible hysteresis loss could make this material promising for low temperature magnetic refrigeration.

SG is thankful to CSIR, New Delhi for granting the fellowship. The authors acknowledge UGC-DAE Consortium for Scientific Research, Indore for providing magnetoresistance (MR) facility. SG also acknowledges CSR, Indore for travel and local hospitality during these measurements.

**Figure Captions:**

Fig. 1. (a) Temperature dependence of dc magnetic susceptibility (left-hand panel) along with the Curie-Weiss fit to the inverse susceptibility data (right-hand panel). (b) Temperature dependence of the magnetic susceptibility (FC) at different fields.

Fig. 2. Field dependence of magnetization isotherms for field up to 50 kOe.

Fig. 3. Temperature dependence of the heat capacity at 0 and 20 kOe field.

Fig. 4.Temperature dependence of isothermal magnetic entropy change (a) at different fields, calculated from magnetization data (b) at 20 kOe, calculated from heat capacity data. The inset in (b) shows temperature dependence of $\Delta T_{ad}$.

Fig. 5. (a) Temperature dependence of the zero field electrical resistivity in ErNiSi. The inset shows the resistivity in 0, 20 and 50 kOe fields. (b) temperature dependence of MR at 20 and 50 kOe fields. The inset shows field dependence of the MR at different temperatures.

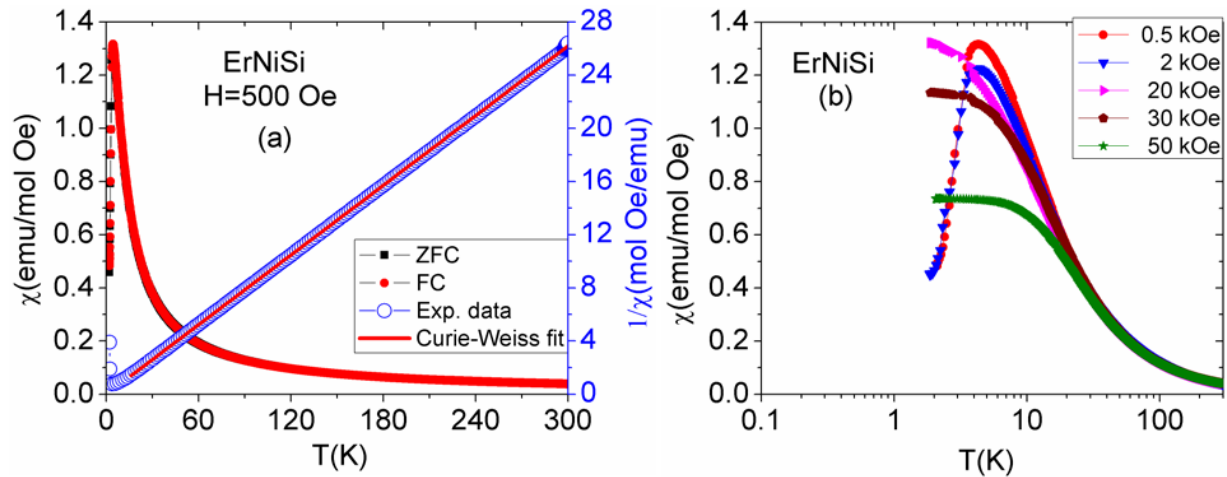

Fig. 1. Sachin Gupta et al.

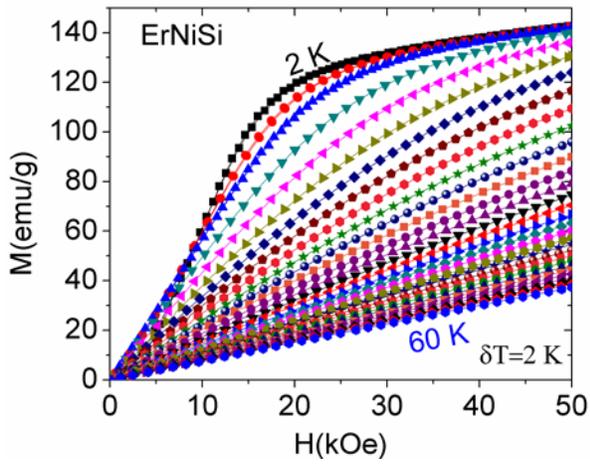
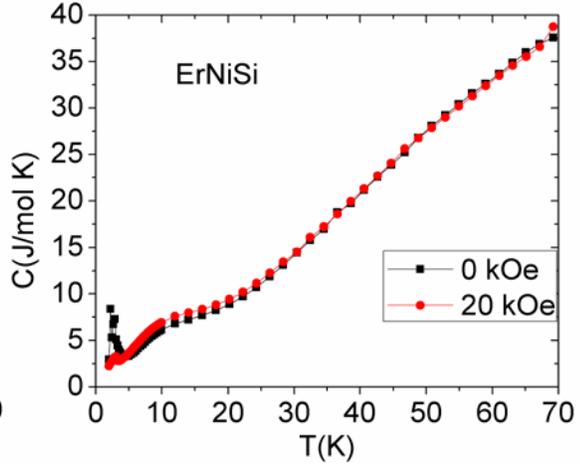

Fig. 2.  Sachin Gupta et al.           Fig. 3. Sachin Gupta et al.

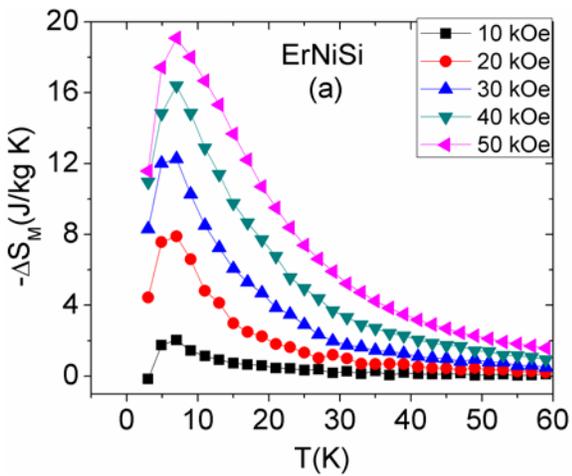
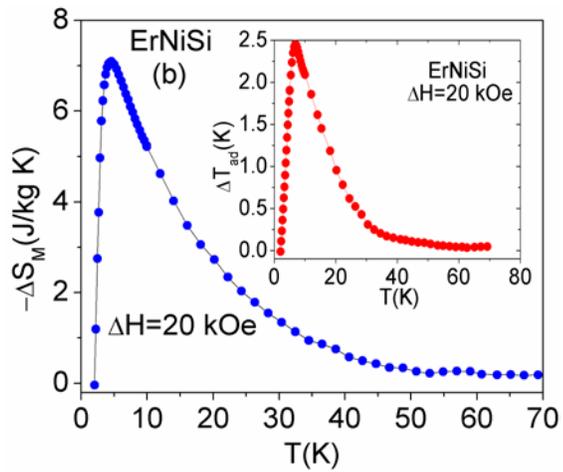

Fig. 4. Sachin Gupta et al.

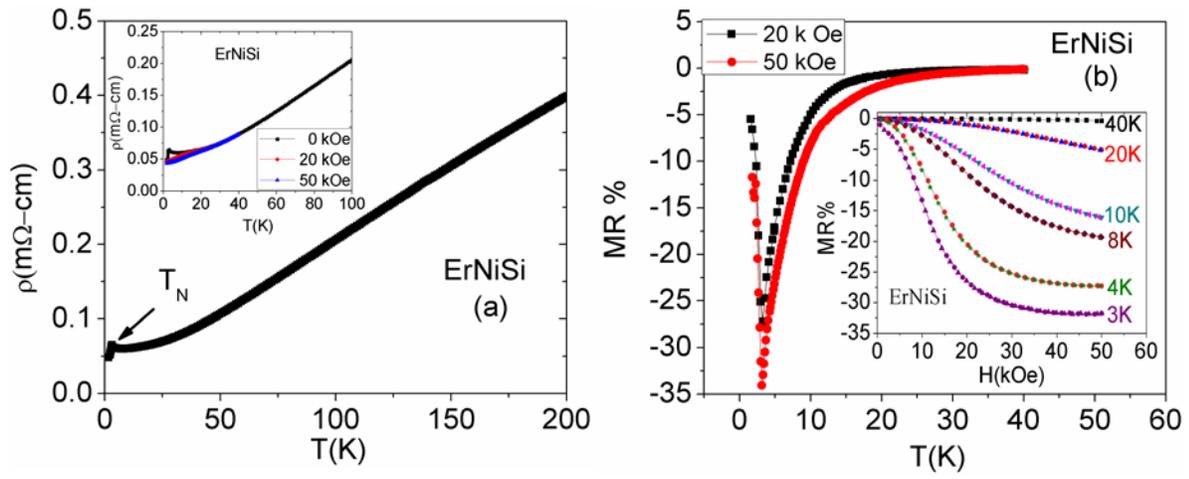

Fig. 5. Sachin Gupta et al.